\documentclass[a4paper,twocolumn,10pt]{IEEEtran}
\pdfoutput=1
\usepackage[pdftex]{graphicx}
\usepackage{amsmath,epsfig,dsfont,url}
\usepackage[nospace]{cite}

\begin{document}

\title{Toward optimal X-ray flux utilization in breast CT}
%
\author{Jakob~H.~J{\o}rgensen$^1$, Per Christian Hansen$^1$,
Emil Y. Sidky$^2$, Ingrid S. Reiser$^2$, and Xiaochuan Pan$^2$
\thanks{
$^1$Technical University of Denmark,
Department of Informatics and Mathematical Modeling,
Richard Petersens Plads, Building 321, 2800 Kgs. Lyngby, Denmark.
$^2$The University of Chicago,
Department of Radiology MC-2026,
5841 S. Maryland Avenue, Chicago IL, 60637. Corresponding author:
Emil Y. Sidky, E-mail: sidky@uchicago.edu}}

%
\maketitle
\thispagestyle{empty}

\begin{abstract}
A realistic computer-simulation of a breast computed tomography (CT)
system and subject is constructed. The model is used to investigate the optimal
number of views for the scan given a fixed total X-ray fluence.  The reconstruction
algorithm is based on accurate solution to a constrained, TV-minimization
problem, which has received much interest recently for sparse-view CT data.
\end{abstract}

\section{Introduction}
\label{sec:intro}
Dose reduction has been a primary concern in diagnostic computed tomography (CT) in recent years \cite{McCollough:09}.
Interest in low intensity X-ray CT is also motivated by the potential to employ CT for screening, where
a large fraction of the population will be exposed to radiation dose and
the majority of subjects will be asymptomatic. This abstract examines the screening application
of breast CT; we simulate breast CT projection data and perform image reconstruction based on constrained,
total-variation (TV) minimization.  The specific question of interest is: given a fixed, total
X-ray flux, what is the optimal number of views to capture in the CT scan?  As the total flux
is fixed, more views implies less photons per view, resulting in a higher noise level per view.
On the other hand, fewer views may not provide enough sampling to recover the underlying
object function.  The optimal balance of these two effects will depend on the imaged subject
and the imaging task. For this reason, we have focused on the breast CT application as a case
study, which also has received much attention in the literature \cite{chen2002cone,kwan2007evaluation,lai2007visibility}.

From the perspective of non-contrast CT, the breast has essentially four gray levels corresponding
to: skin, fat, fibro-glandular or malignant tissue, and calcification. In designing the CT system, physical
properties of the subject that are important are the complexity of the fibro-glandular tissue, which
could be the limiting factor in determining the minimum number of views in the scan, and micro-calcifications
and tumor spiculations, which challenge the resolution of the system.

The image reconstruction algorithm, investigated here, is based on accurate solution of constrained,
TV-minimization.  Constrained, TV-minimization is reconstruction by solving an optimization problem
suggested in the compressive sensing (CS) community for taking advantage of sparsity of the subject's
gradient magnitude \cite{candes2006robust,candes2008introduction}.
Various algorithms based on TV-minimization have been investigated
for sparse-view CT data
\cite{SidkyTV:06,song2007sparseness,chen2008prior,sidky2008image,Bergner:10,Choi:10,Bian:10},
but we have also recently begun investigating TV-minimization
for many-view CT with a low X-ray intensity. While the emphasis in many of these
works has been algorithm efficiency, the aim here is different in that we seek accurate solution
to TV-minimization in order to simplify the trade-off study. With accurate solution of TV-minimization,
the resulting image can be regarded as a function of only the parameters of the optimization
problem, removing the additional variability inherent in inaccurate but efficient TV-minimization
solvers. The actual solver used here employs an accelerated gradient-descent algorithm which is
described in an accompanying abstract and in Ref. \cite{Jensen,Jorgensen3D}. This solver allows us to investigate
the behavior of the solution to constrained, TV-minimization as the number of projections is varied
at fixed total flux. As this is a preliminary study, the evaluation is based upon visual inspection
of images obtained with
a realistic computer-phantom and a CT data model incorporating physics of the low-intensity
scan. Section \ref{sec:model} describes the system and subject model in detail; Sec. \ref{sec:recon}
briefly describes the reconstruction algorithm; and Sec. \ref{sec:results} presents indicative 
results of the sampling/noise trade-off study for breast CT.

\section{Breast CT model}
\label{sec:model}

We model the salient features of a low intensity X-ray CT system and a breast subject
to gain an understanding of the trade-off between noise-per-projection and number-of-projections.

\subsection{phantom}

The breast phantom has four components: skin, fat, fibro-glandular tissue and micro-calcifications.
The latter two components are the most relevant and are now described in detail.
We refer all gray values to that of fat, which is taken to be 1.0.
The skin gray level is set to 1.15.\\
{\it Fibro-glandular tissue:} The gray value is set to 1.1.
The pattern of this tissue is generated by a power law noise model described
in Ref. \cite{reiser2010task}. The complexity of this tissue's attenuation map
is similar to what one could find in a breast CT slice.
For the present study, the background fibro-glandular tissue, fat and skin are
represented with as a 1024x1024 digital phantom, from which projections are computed.
The reason for doing so, is that we want to isolate the issue of structural complexity
of the background, while removing potential ambiguity of
projection model mismatch. \\
{\it Micro-calcifications:} 5 small ellipses with attenuation values ranging from 1.8 to 2.1.
In this
case, the ellipse projections are generated from a continuous ellipse model, and
unlike the rest of the phantom, these projections are not consistent
with the digital projection system matrix.  For these structures, object pixelization
is a highly unrealistic model because of their small size; hence we employ the
continuous model to generate their projection data.

\begin{figure}
\begin{minipage}[b]{0.99\linewidth}
\centering
\centerline{\includegraphics[width=9cm,clip=TRUE]{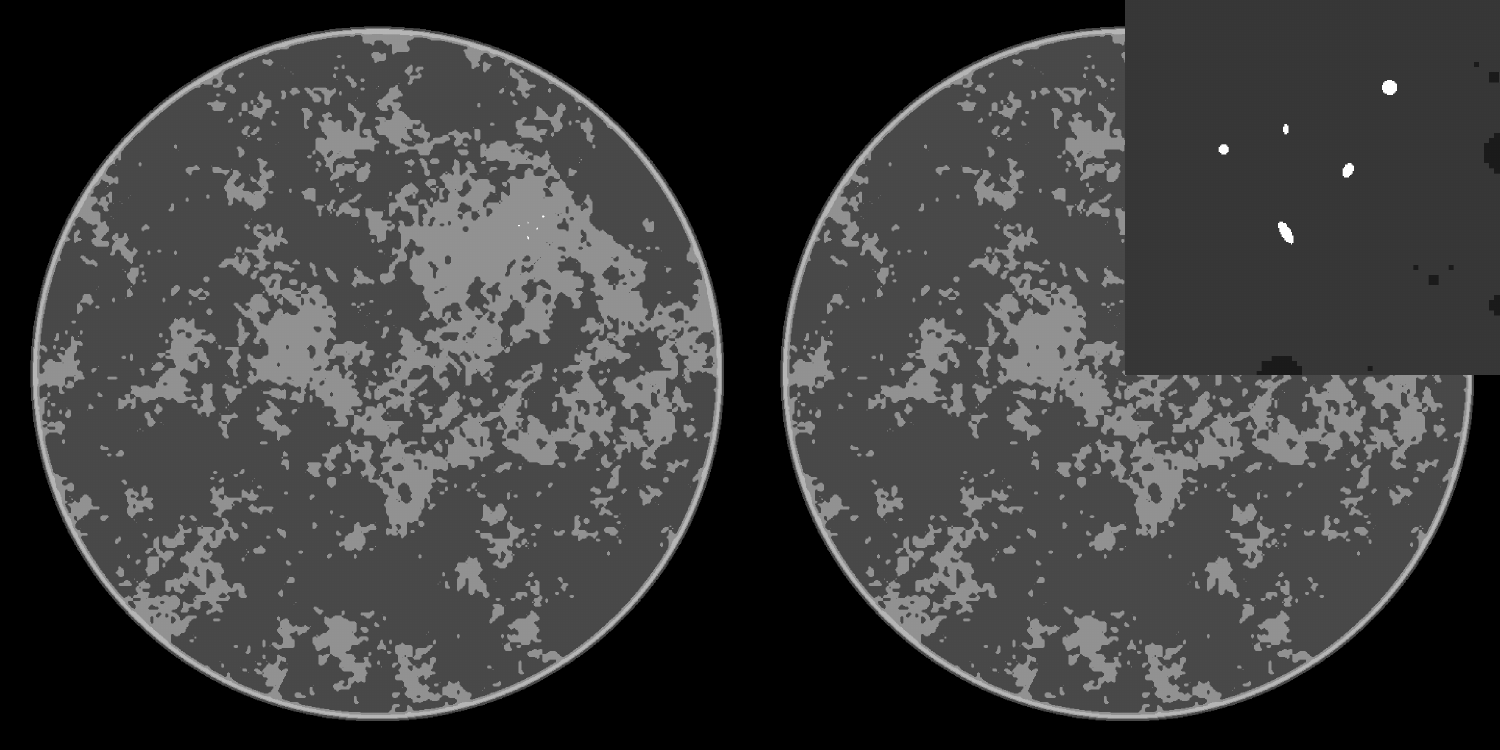}}
\end{minipage}
\caption{Left: complete breast phantom shown in a gray scale window [0.9,1.25].
Right: same phantom with a blown-up inlay of 7.5x7.5 mm$^2$ ROI containing the
micro-calcifications. The ROI grayscale window is [0.9,1.8].
All image reconstruction results are shown in this format.
\label{fig:breastPhantom}}
\end{figure}

The complete phantom along with a blow-up of a region of interest (ROI)
containing the micro-calcifications is shown in Fig. \ref{fig:breastPhantom}.
The complexity of background is apparent, and although the phantom is indeed piece-wise constant,
the gradient magnitude has ~55,000
non-zero values due to the structure complexity.
This number is relevant for the CS argument on the accuracy of
TV-minimization. While there has been no analysis of CS recovery for CT-based
system matrices, one can expect that at least twice as many samples as non-zero
elements in the gradient magnitude will be needed for accurate image
reconstruction with TV-minimization under noiseless conditions.

\subsection{data model}
\label{sec:datamodel}

As the primary goal of this study is to investigate a noise trade-off, the
CT model includes a random component modeling the detection of finite numbers of X-ray quanta.
The process of generating the simulated CT data starts with computing a noiseless
sinogram:
\begin{equation}
g_j = \int_{L_j} d\ell f_\text{digital}[\vec{r}(\ell)] + f_{\mu \text{calc}}[\vec{r}(\ell)],
\end{equation}
where $g_j$ is the $j$th line integral of the phantom over the ray $L_j$ with
the index $j$ running from 1 to $N_\text{data}$; $N_\text{data}$ is the product 
of the number of projections and the number of detector bins per projection;
and $f_\text{digital}[\vec{r}(\ell)]$ and $f_{\mu \text{calc}}[\vec{r}(\ell)]$
represent the digital and continuous components of the phantom, respectively.
The measurements $g_j$ are used for the noiseless reconstructions.

In order to include a random element to the data, which depends on $N_\text{data}$
in a fairly realistic way, we compute a mean photon number per detector
bin based on $g_j$ and
a total photon intensity of the scan:
\begin{equation}
\notag
n^\text{(mean)}_j = \frac{N_\text{photon}}{N_\text{data}}\exp (- g_j),
\end{equation}
where $N_\text{photon}$ is the total number of photons in the scan and is here
selected to be a value typical of mammography.
Note that the model the scale factor will cause the mean number of photons
per bin to decrease as the number of ray measurements increases.
From $n^\text{(mean)}_j$, a realization $\mathbf{n}_j$ is selected from a Gaussian distribution,
using $n^\text{(mean)}_j$ as the mean and variance. This Gaussian distribution closely
models a Poisson distribution for large $n^\text{(mean)}_j$. Finally, the photon number
noise realization is converted back to a realization of a set of line integrals:
\begin{equation}
\notag
\mathbf{g}_j = - \ln \left( \frac{N_\text{data}}{N_\text{photon}} \mathbf{n}_j \right).
\end{equation}
It is this data set which will be used for the noisy reconstructions below.
While this model incorporates the basic idea of the noise-level trade-off, there
are still limitations of the study. The incident intensity on each detector bin is
assumed to be the same; no correlation with neighboring bins is considered; electronic
noise in the detector is not accounted for; and reconstructions are performed from
a single realization as opposed to an ensemble of realizations.

\section{Image reconstruction by constrained TV-minimization}
\label{sec:recon}

In order to perform the image reconstruction, we employ 
CS-motivated, constrained, TV-minimization:
\begin{equation}
\label{ctv}
\vec{f}^* = \text{argmin} \|\vec{f} \|_\text{TV} \text{ subject to } |X\vec{f} - \vec{ \mathbf{g} } |^2 \le
\epsilon^2 \text{ and } \vec{f} \ge 0,
\end{equation}
where the norm $ \| \cdot  \|_\text{TV}$ is the sum over the gradient magnitude of the image;
the system matrix $X$ represents discrete projection converting the image estimate $\vec{f}$
to a projection estimate $\vec{g}$; $\epsilon$ is a data error tolerance parameter controlling how
closely the image estimate is constrained to agree with the available data; and the last
constraint enforces non-negativity of the image.
This optimization problem has served to aid in designing many new image reconstruction algorithms
for CT. As the CT application is quite challenging, most of these algorithms do
not yield the solution $\vec{f}^*(\epsilon)$ of Eq. (\ref{ctv}), which should only depend on
$\epsilon$ once the CT system parameters are fixed. As a result, these algorithms yield images which
also depend on algorithm parameters. This is not necessarily a bad thing, but it becomes difficult
to survey the effectiveness of Eq. (\ref{ctv}) for various CT applications.

In applied mathematics, motivated by CS, there has been much effort in developing accurate solvers
to Eq. (\ref{ctv}), but few of these solvers can be applied to systems as large as those encountered
in CT. To address this issue, we have been investigating means of accelerating gradient methods, which
can be implemented for systems on the scale typical of CT. The proposed set of algorithms are described
in detail in an accompanying submission to the meeting \cite{Jorgensen3D}. We do not discuss the algorithm
here, but we point out that the optimization problem solved is modified, but equivalent to Eq. (\ref{ctv}):
\begin{equation}
\label{uctv}
\vec{f}^* = \text{argmin } \alpha \|\vec{f} \|_\text{TV} + |X\vec{f} - \vec{ \mathbf{g} } |^2 
\text{ subject to } \vec{f} \ge 0,
\end{equation}
where the data error term has been included in the objective function, leaving only positivity as a constraint.
The penalty parameter $\alpha$ replaces the role of $\epsilon$ above. We use the accelerated gradient algorithm
to solve Eq. (\ref{uctv}) to a numerical accuracy greater than what would be visible in the images; thus,
we describe the following resulting images as solutions to this optimization problem. To make the connection
with the Eq. (\ref{ctv}) is straight-forward; the corresponding $\epsilon$ to a given $\alpha$ is found
by computing $|X \vec{f}^* - \vec{ \mathbf{g} } |$ where $\vec{f}^*$ is found from Eq. (\ref{uctv}).

\section{Results}
\label{sec:results}

\begin{figure}
\begin{minipage}[b]{0.99\linewidth}
\centering
\centerline{\includegraphics[width=9cm,clip=TRUE]{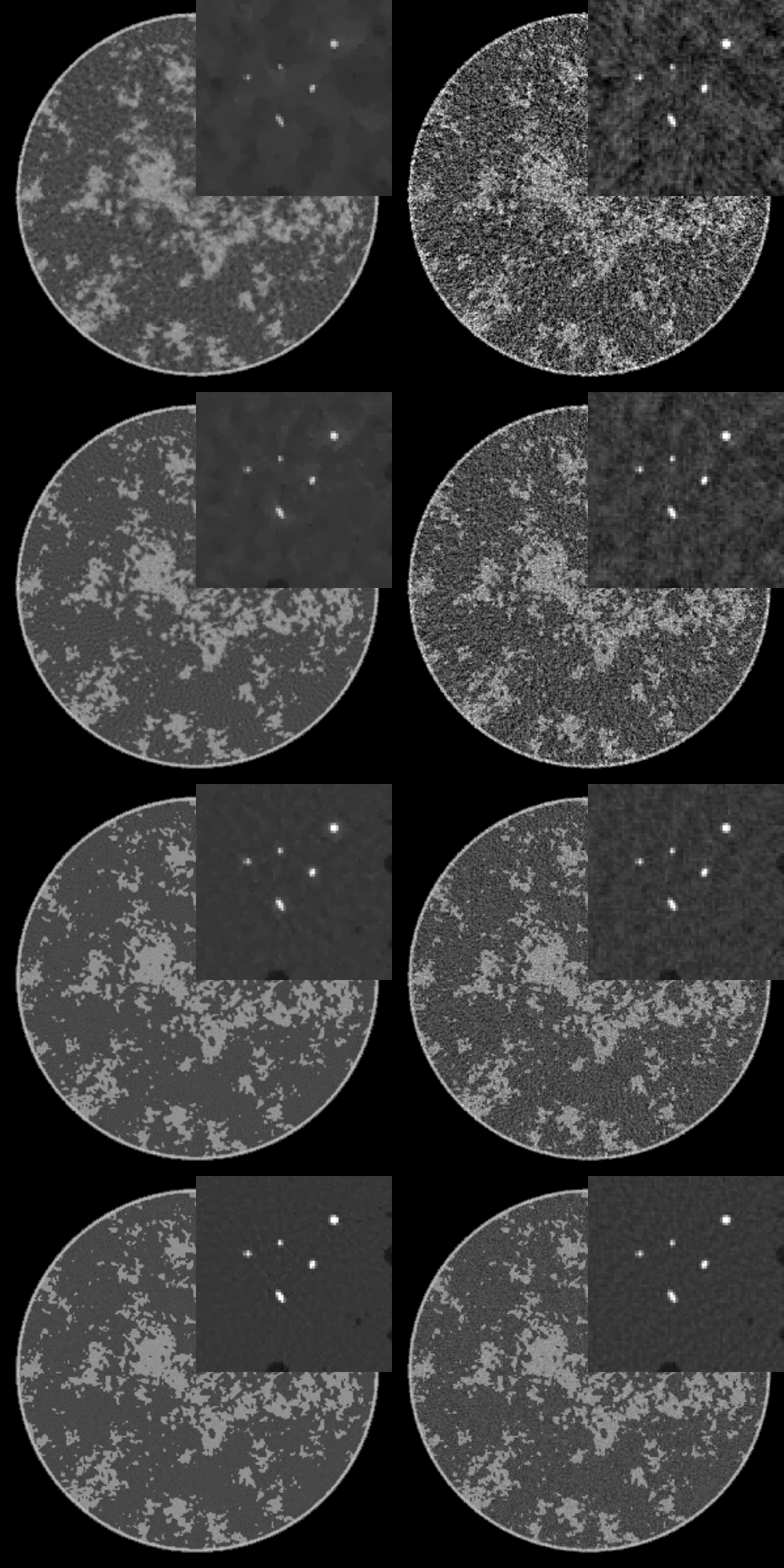}}
\end{minipage}
\caption{Left column: images reconstructed by TV-minimization. Right column:
images reconstructed by FBP. The data do not include noise, and the number
of views are 64, 128, 256, and 512 going from top to bottom.
\label{fig:noiseless}}
\end{figure}

For this initial survey of a breast CT simulation, we show two main sets of results. The first
set of images are reconstructed from noiseless data for different numbers of views. The idea
is to see how well TV-minimization performs in recovering the complex breast phantom under ideal
conditions. The second set of images includes noise at a fixed exposure, and as described in
Sec.~\ref{sec:datamodel}, the noise-level per projection increases with the the number of projections.

All reconstructions are performed on a 1024x1024 grid with 100 micron pixel widths. The simulated
fan-beam geometry has an 80 cm source to detector distance with a circular source trajectory of radius
60 cm. The detector is modeled as having 1024 detector bins, and there is no truncation in the projection
data. 

\begin{figure}
\begin{minipage}[b]{0.99\linewidth}
\centering
\centerline{\includegraphics[width=9cm,clip=TRUE]{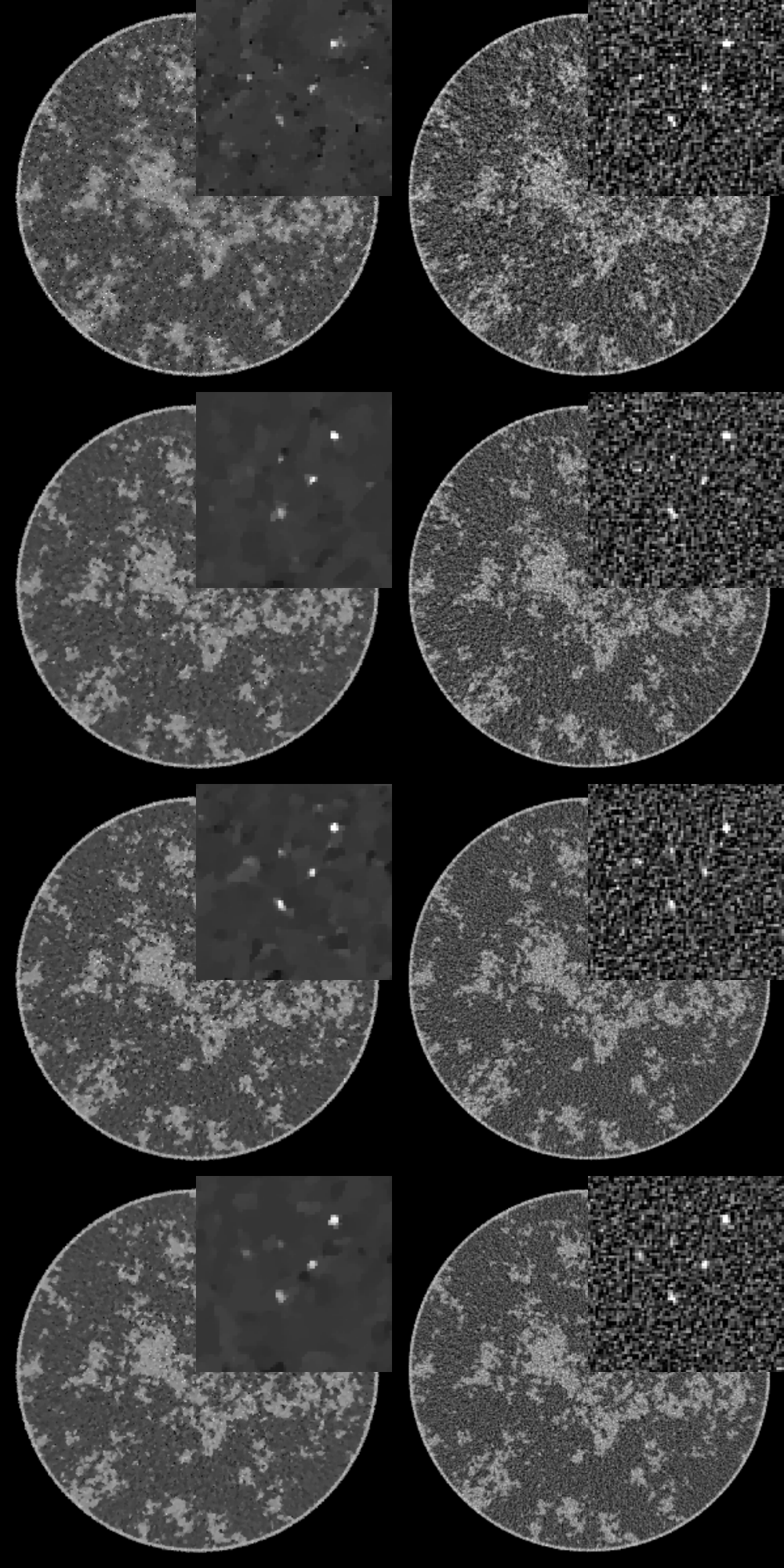}}
\end{minipage}
\caption{Same as Fig. \ref{fig:noiseless} except the noise
model discussed in Sec. \ref{sec:datamodel} is included.
\label{fig:noisy}}
\end{figure}

\subsection{image reconstruction from noiseless data}

In Fig.~\ref{fig:noiseless}, we show images reconstructed from 64 to 512 projections for
both TV-minimization and filtered back-projection (FBP). For TV-minimization in this study
we set $\alpha=10^{-6}$, which corresponds to a very tight data constraint. As noted above
the sparsity of the gradient magnitude is on the order of 50,000. Accordingly, from
CS-based arguments, one could only expect to start to achieve accurate reconstruction
when the number of measured line integrals exceeds 100,000, which in this case means
100 projections. An important part of CS theory
deals with computing the factor between the sparsity level and necessary number of measurements
for accurate recovery. This factor is unknown for TV-minimization applied to the X-ray transform,
but we can see from the reconstructions that the accuracy is greatly improved in going from
128 views to 256 views. There is still a perceptible improvement in the image
recovery in going to 512 views, which still represents an under-determined
system despite the fact that 512 views is normally not thought of as a sparse-view
data set. Again, it is the complexity of the phantom which is responsible for this
behavior.
The accompanying FBP results give an indication on the ill-posedness
of reconstruction from the various configurations with different numbers of projections.

The results for the micro-calcification ROI are interesting
in that this particular feature of the image is recovered for all data sets down to the 64-projection
data set.  This is not too surprising because the micro-calcifications are certainly sparse
in the gradient magnitude. But this result emphasizes that the success of an image reconstruction
algorithm depends also on the imaging task and the subject.

For the larger goal of determining the optimal number
of views, it is clear that "structure noise" --
artifacts due to the complex object function--
can play a significant role for this breast phantom.

\subsection{image reconstruction from noisy data}

\begin{figure}
\begin{minipage}[b]{0.99\linewidth}
\centering
\centerline{\includegraphics[width=9cm,clip=TRUE]{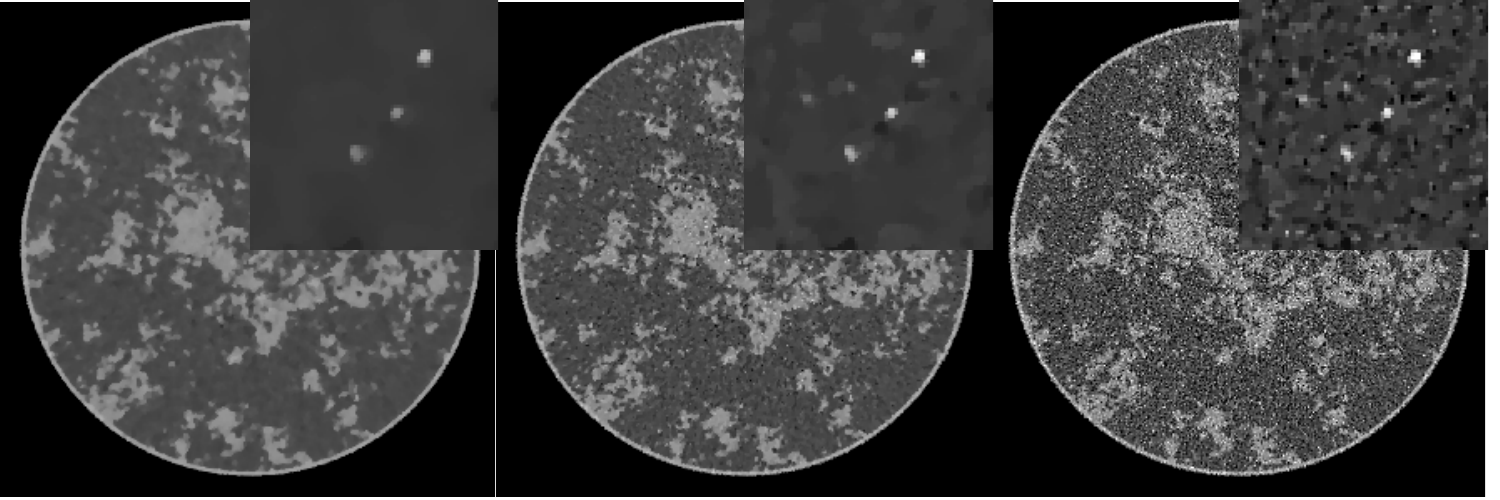}}
\end{minipage}
\caption{Images for 512-view, noisy projection data obtained with TV-minimization
for (left) $\alpha=1 \times 10^{-3}$, (middle) $\alpha=5 \times 10^{-4}$, and (right)
$\alpha=2 \times 10^{-4}$.
\label{fig:alpha}}
\end{figure}

For the noise studies, we again investigate data sets with the view number varying between
64 and 512. For these reconstructions, $\alpha$ is also varied between
$1.\times 10^{-6}$ and
$5. \times 10^{-4}$. In Fig.~\ref{fig:noisy}, we show the TV-minimization images compared
with FBP, as a reference. The optimal values of $\alpha$ for each TV-minimization image
is chosen by visual inspection. The FBP fill images are smoothed by convolving 
with a Gaussian distribution of width 140 microns (chosen by visual inspection), and
the ROI images are unregularized. While it is not too surprising that the FBP image
quality appears to increase with projection number, it is
somewhat surprising that the same trend is apparent for image reconstruction
by TV-minimization. The 512-view data set seems to yield, visually, the optimal
result in that the ROI appears to have the least amount of artifacts. While most
of the micro-calcifications are visible in each reconstruction, the artifacts and
noise texture in
the sparse-view images can be distracting and mistaken for additional
micro-calcifications. It seems that
the increased noise-level per view impacts the reconstruction less than artifacts
due to insufficient sampling. That we obtain this result with a CS algorithm is
interesting, and warrants further investigation with more rigorous and quantitative
evaluation.

To appreciate the impact of $\alpha$, we focus on the 512-view data set
and display images in Fig. \ref{fig:alpha} for three cases. Small $\alpha$
corresponds to a tight data constraint, resulting in salt-and-pepper
noise in the image due to the high noise-level of the data. Increasing
$\alpha$ reduces the image noise and eventually removes small structures.

\section{conclusion}

We have performed a preliminary investigation of a fixed
X-ray exposure trade-off between number-of-views and noise-level
per view for a simulation of a breast CT system. This investigation
employed a CS image reconstruction algorithm which should favor
sparse-view data. Moreover, the simulated data are generated from
a digital projection matched with the projector used in the image
reconstruction algorithm -- another factor that should favor sparse-view
data. Despite this, the complexity of the subject overrides these
points and it appears that the largest number of views, in the study, yields
visually the optimal reconstructed images.  When other physical factors
are included in the data model, for example, partial volume averaging and X-ray
beam polychromaticity, one can expect that this same conclusion
will hold.

Extensions to the image reconstruction algorithm will address better
noise modeling. One can expect an improvement in image quality by
employing a weighted, quadratic data error term derived from a realistic
CT noise model. As for CS-motivated image reconstruction, the breast CT
system may benefit from exploiting other forms of sparsity.

\section{Acknowledgments}
This work is part of the project CSI: Computational \hbox{Science} in Imaging,
supported by grant 274-07-0065 from the Danish Research Council for Technology and Production Sciences.
E.Y.S.  and X.P. were supported in part by NIH
R01 Grant Nos. CA120540 and EB000225.
I.S.R. was supported in part by
NIH Grant Nos. R33 CA109963 and R21 EB8801.
The contents of this article are solely the responsibility of
the authors and do not necessarily represent the official
views of the National Institutes of Health.

\bibliographystyle{ieeebib}
\bibliography{lowdose.bib}
\end{document}